\documentclass[]{emulateapj}
\usepackage{natbib}
\usepackage{threeparttable}
\usepackage{amsmath}
\usepackage{amssymb}
\usepackage{graphicx}
\usepackage{subfigure}
\usepackage{natbib}
\newcommand{\msun}{\ensuremath{M_\odot}}

\newcommand{\rsun}{\ensuremath{R_\odot}}

\newcommand{\ptran}{\mbox{$P_{\rm tran}$}}

\shortauthors{Imara \& Di Stefano}

\begin{document}

\title{Searching for Exoplanets Around X-Ray Binaries with Accreting White Dwarfs, Neutron Stars, and Black Holes}

\author{Nia Imara \& Rosanne Di Stefano}
\affil{Harvard-Smithsonian Center for Astrophysics, 60 Garden Street, Cambridge, MA 02138}

\email{nimara@cfa.harvard.edu}

\begin{abstract}
We recommend that the search for exoplanets around binary stars be extended 
to include X-ray binaries (XRBs) in which the accretor is a white dwarf, neutron star, 
or black hole.  
We present  a novel idea for detecting planets bound to such mass 
transfer binaries: 
we propose that the X-ray light curves of these binaries be inspected 
for signatures of transiting planets. X-ray transits may be the only way to 
detect planets around some systems, while providing a complementary approach 
to optical and/or radio observations in others.  
Any planets associated with XRBs must be in stable orbits.
We  consider the range of allowable separations and find that 
orbital periods can be hours or longer, while transit 
durations extend upward from about a minute for Earth-radius
planets, to hours for Jupiter-radius planets.
The search for planets around XRBs could begin at 
once with existing X-ray observations of these systems.  
If and when a planet is detected around an X-ray binary, 
the size and mass of the planet may be readily measured, and it
may also be possible to study the transmission and absorption of
X-rays through its atmosphere.
Finally, a noteworthy application of our proposal is that the same technique could 
be used to search for signals from extraterrestrial intelligence.  
If an advanced exocivilization placed a Dyson sphere or similar structure in orbit 
around the accretor of an XRB in order to capture energy, 
such an artificial structure might cause detectable transits in the X-ray light curve.
\end{abstract}

\keywords{planets and satellites: detection --- planetary systems --- binaries: eclipsing --- white dwarfs --- X-rays: binaries --- novae, cataclysmic variables }

\section{Introduction}

X-ray binaries (XRBs) are gravitationally interacting binaries 
in which a stellar remnant accretes matter from its companion. The term XRB is often used for 
binaries in which the accretor is a neutron star or black hole, and here we will 
apply it as well
to binaries in which the accretor is a white dwarf. 
We include both cataclysmic variables (CVs),
in which the white dwarf accretes from a close low-mass companion,
and symbiotics, in which the donor is a giant.
We consider the possibility that some XRBs host planetary systems, 
and we present a method to discover planets orbiting within or around such 
interacting binaries. 

\subsection{Planets in Mass-Transfer Binaries: Feasibility}

Planets in binaries may be common.  At the time of this writing, roughly 70 examples are known of planets orbiting one member of a binary \citep[e.g.,][]{Butler_1997, Cochran_1997, Roell_2012}, and more than a dozen cases are known in which the planet orbits both members  \citep[e.g.,][]{Thorsett_1999, Doyle_2011, Hinse_2015, Orosz_2012a, Orosz_2012b}.  
Following the terminology of \citet{Dvorak_1986}, these two types of systems are S-type and P-type, respectively.  The first known P-type system consists of a planet orbiting a  pulsar and a white dwarf and was detected from observed changes in the travel time of the radio pulses \citep{Thorsett_1999}.  This system passed through an epoch as an X-ray binary, during which
 the giant progenitor of today's white dwarf transferred mass to the neutron star, spinning it up  Here we consider how to search for planets in active mass-transfer systems. 

Planets have also been discovered in P-type orbits
around binaries which {\sl will become} CVs
during their next evolutionary phase (e.g., HW Vir, Lee et al. 2009; HW Vir, Beuermann et al. 2012; DP Leo, Qian et al. 2010; NN Ser, Beuermann et al. 2010).  These binaries are referred to as {\sl pre-CVs} or {\sl post-common-envelope} binaries.     Pre-CVs contain a white dwarf which is the remnant of a giant that filled its Roche lobe under  circumstances inconsistent with dynamically stable mass transfer. The giant's envelope was transformed into a common envelope within which the white dwarf and its stellar companion spiraled closer to each other, but did not merge.  Eventually, the white dwarf's companion will fill its Roche lobe, and the binary will become a CV. If the planet is retained, it will be in orbit with an XRB. 

Since planets have been found in P-type orbits both prior to and after the binary is an XRB, it is clearly important to consider planets during the XRB phase.
Planets in S-type orbits around accreting compact objects or companion donor stars may also be present, especially in wider binaries, where mass transfer may proceed through winds. Those planets not destroyed by mass transfer may transit the compact object and portions of its accretion disk, potentially causing detectable X-ray eclipses. The same considerations apply, whether the accretor is a white dwarf, neutron star, or black hole.

%%%%%%%%%%%%%%%%%%%%%%%% FIGURE 1 %%%%%%%%%%%%%%%%%%%%%%%%%%%%%%%%%%
\begin{figure*}[ht]
\centering
\includegraphics[width=5in]{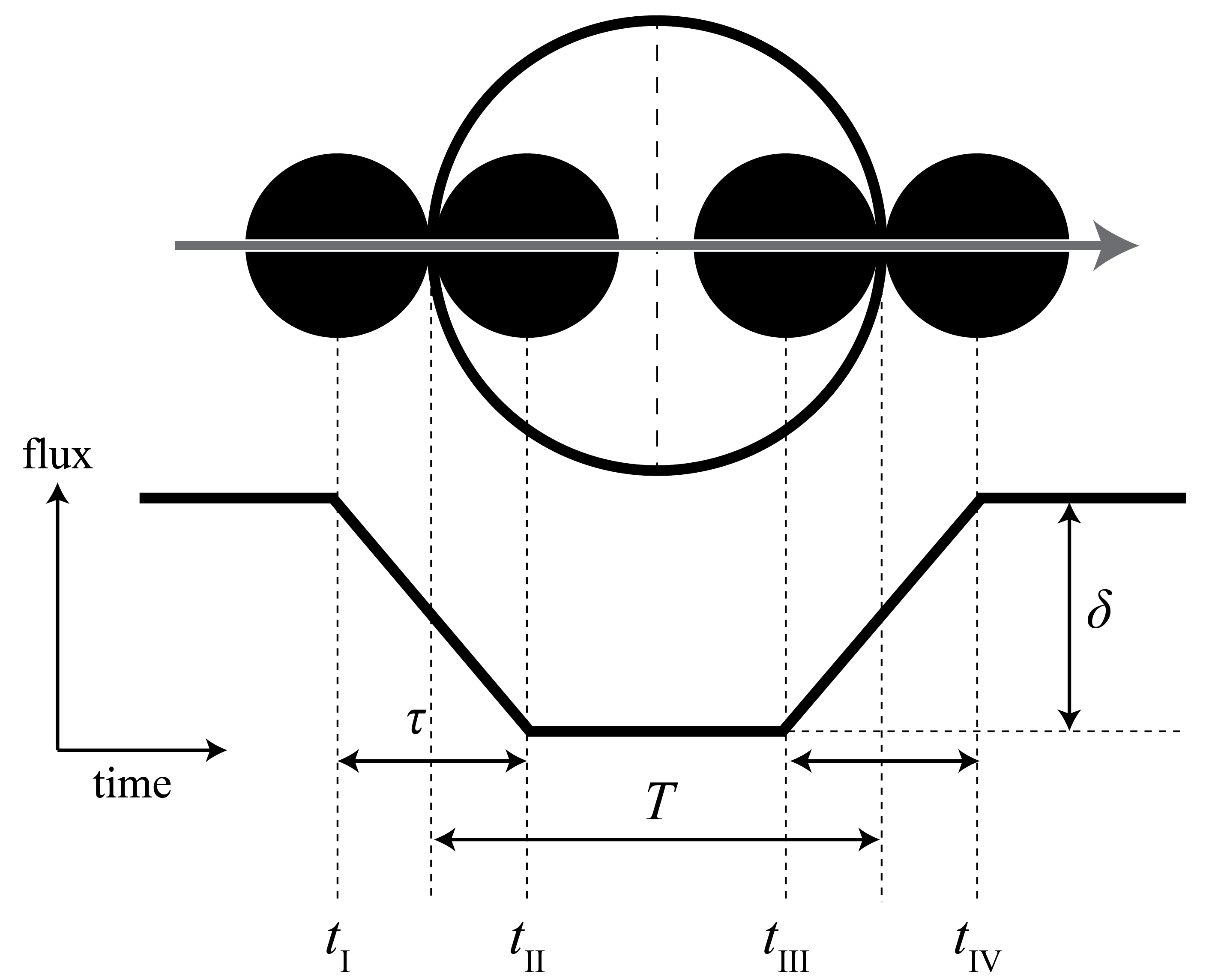}
\caption{Illustration of a transit.  The eclipsed compact object is represented by the large circle, and the transiting planet is represented by the black circles at four contact points, at times $t_I$, $t_{II}$, $t_{III}$, and $t_{IV}$.  A model light curve is shown with labels for the transit duration, $T$, and the ingress/egress duration, $\tau$.   \label{fig:cartoon}}
\end{figure*}
%%%%%%%%%%%%%%%%%%%%%%%%%%%%%%%%%%%%%%%%%%%%%%%%%%%%%%%%%%%%%%%%%%%%

\subsection{The Possibility of Detection Through X-ray Transits}\label{sec:possibility}
X-ray light curves provide ideal opportunities to search for transits, because the sizes of the X-ray emitting regions in many XRBs are so small that the passage of a planet produces a sharply-defined total eclipse.  For example, X-ray emission from black holes can be dominated by a soft ($\sim$ keV) component emanating primarily from the inner portion of an accretion disk, which may have an extent of only tens or hundreds of kilometers \citep[e.g.,][]{Remillard_2006}. X-rays from neutron stars may also emerge from very small regions. In certain states, emission from a white dwarf emanates from an area comparable to its surface area, so that an Earth-size planet could cause a deep eclipse, and somewhat larger planets could produce total eclipses. This would apply to a white dwarf in a supersoft state \citep[e.g.,][]{VanDenHeuvel_1992}, or post-nova, and also to some magnetic CVs \citep[e.g.,][]{Nucita_2009}.    

The question of detectability hinges on whether the number of photons 
missing due to the eclipse is large enough that the deficit and its 
duration can be reliably measured. If reliable detection cannot be made 
during a single transit, then the signal could possibly be identified 
through repeated transits, if the orbital period is short enough.
The challenges are similar to those that have been successfully met by analyses of data from the {\sl Kepler} mission,
which has identified Earth-size and even
smaller planets orbiting Sun-like stars. The X-ray data analyses we advocate
would take place in a complementary part of the discovery parameter space:
the X-ray transits are deeper, in some cases total eclipses; the background
is small or nonexistent, eliminating problems of blending with light from
nearby stars; and the numbers of X-ray photons are small.
The paucity of X-ray photons is a new challenge which provides a lower bound on
the baseline count rate relative to the eclipse time scale and the total exposure time (see \S 4).  Nevertheless, if planets orbit within or around XRBs, then detectability is possible for many systems.

\subsection{Complementary Methods}
X-ray binaries are also copious emitters of optical light.  In cases in which the optical emission, like the X-ray emission, emanates from a small region, optical transits may also be detected. HT Cas and OY Car are examples of CVs in which
both optical and X-ray eclipses have been studied \citep[][]{Wood_1989, Horne_1991, Wheatley_2003, Thorstensen_2008, Nucita_2009}.  In many XRBs, however, the sources of optical emission can be complex and extended. The donor star and outer regions of the accretion disk can both contribute to or dominate the optical emission, diluting the effects of stellar or planetary transits.  In these cases, X-rays may provide the best route to planet detection.
Moreover, the detection of transits in two widely separated 
wavebands increases confidence in the interpretation that the dips are
related to the passage of a solid body. In addition, different
wavebands reveal complementary aspects of the emission during mass transfer.
Periodic pulsations, like those emanating from millisecond pulsars, 
 can reveal the presence of planets through radio timing measurements. 
As with optical observations, combining information across
wavebands is an exciting prospect.

\subsection{Plan of this Paper} 
We demonstrate that planets 
may produce detectable X-ray eclipses.  We consider two extremes of planet 
masses and sizes, from Earth- to Jupiter-sized.  Our main goals are 
(1) to examine the parameter space of stable orbits for planets in XRBs; 
(2) to study the expected light curve characteristics; and 
(3) to consider which XRBs would likely be most profitable to study with our approach.  

In Section \ref{sec:stability}, we present the equations for 
investigating the stability limits of planets orbiting XRBs, 
with a specific focus on CVs.  We provide examples of theoretical 
transiting light curves in Section \ref{sec:light_curves}. We
address the issue of detectability and consider  
the properties of XRBs that 
are the most promising targets for light curve analysis 
in Section \ref{sec:discussion}.
In Section \ref{sec:conclusion} we summarize our conclusions
and discuss the implications of our work.

\section{Orbital Stability}\label{sec:stability}
Our goal in this section is to explore a realistic range within the  
parameter space of stable orbits for planets in XRBs. For simplicity, we restrict our considerations to single planets in circular, coplanar orbits.  An illustration of a planet orbiting and transiting the compact accretor of an XRB is shown in Figure \ref{fig:cartoon}.

Stability in triple systems generally requires a hierarchical 
structure, with the outer orbit being at least a few times
larger than the inner orbit. Thus, for P-type (or, circumbinary) orbits, there is a critical value, $a_{c,{\rm out}}$, for the semimajor axis of the planetary orbit in terms of the 
separation $a_b$ between the components of the binary.  \citet{Holman_1999} determined an expression for $a_{c,{\rm out}}$ 
by conducting numerical simulations of test 
particles moving in the gravitational field of binaries:
\begin{equation}\label{eq:outer}
\begin{split}
\frac{a_{c, \rm out}}{a_b} =&~(1.60\pm 0.04) + (5.10\pm 0.05)e       \\
                          & + (-2.22\pm 0.11)e^2 + (4.12\pm0.09)\mu \\
                          & +(-4.27\pm0.17)e\mu + (-5.09\pm 0.11)\mu^2 \\
                          & + (4.61\pm 0.36) e^2 \mu^2.
\end{split}
\end{equation}
Here, $e$ is the orbital eccentricity, which will generally be zero in the mass transfer
binaries we consider. The
binary mass ratio is $\mu=M_2/(M_1+M_2)$, where 
$M_1$ is the mass of primary, and $M_2$ is the mass of its stellar companion.  
In our case, without loss of generality, we define the primary to be the 
compact accretor.  This equation applies over the range $0.1\le\mu\le 0.9$ \citep{Holman_1999}. For $a>a_{c, \rm}$ the orbit is stable.

\begin{table}\centering
\begin{center}
\begin{tabular}{lc}
\multicolumn{2}{c}{Table 1: Variables used in this paper.}\\
\tableline\tableline
Symbol   & Variable definition \\  
\tableline
$M_1~(M_2)$  & Mass of primary (donor) star  \\
$R_1~(R_2)$  & Radius of primary (donor) star   \\
$M_p, R_p, P_p$   & Planet mass, radius, and period    \\
$\mu$        & Mass ratio: $\mu = M_2/(M_1 + M_2)$   \\
$q$          &  $q=M_2/M_1$                          \\
$k$          & Planet-to-primary size ratio: $k=R_p/R_1$ \\
$a_b$        & Semi-major axis of binary   \\
$a_c$        & Critical semi-major axis of planetary orbit   \\
$P$          & Orbital period of planetary orbit \\
$e, b$       & Eccentricity, impact parameter   \\
$i_p~(i_b)$  & Inclination angle of planetary (binary) orbit \\
$\delta i$   & Mutual inclination of planetary and binary orbits  \\
$R_T$        & Tidal disruption radius        \\
$T~(T_{\rm tot})$  & Full (total) duration time of transit  \\
$\tau$       & Ingress/egress time of transit  \\
$\delta$     & Transit depth                   \\
$P_{\rm tran}$& Transit probability              \\
\tableline
\end{tabular}
\label{table1}
\end{center}
\end{table}

It is also possible for the planet to be located within the binary
orbit. In this case, \citet{Holman_1999} find that the critical orbital radius of an S-type planet, $a_{c, \rm in},$ is  
\begin{equation}\label{eq:inner}
\begin{split}
\frac{a_{c, \rm in}}{a_b} =&~(0.464\pm0.006) + (-0.380\pm 0.010)\mu   \\
                         & +(-0.631\pm 0.034)e + (0.586\pm 0.061) e \mu \\
                         & +(0.150\pm0.041) e^2 + (-0.198\pm 0.074) \mu e^2.
\end{split}
\end{equation}
Equation  (\ref{eq:outer}) is valid to within 3--6\% over the range  $0.0\le e \le 0.7$, and equation (\ref{eq:inner})  
is valid to within 4--11\% for $0.0\le e \le 0.8$.  The variables used in the above equations and the rest of this paper are summarized in Table 1.

The specific values we derive for the innermost and outermost stable orbits
are not expected to represent exact results for all planets in binaries,
because some of the input assumptions may not be satisfied.  For example,
the orbits may not be coplanar, there may be a distant third star, 
multiple planets, or the process of mass transfer may alter the results somewhat. Nevertheless,
the general result should be robust: the planetary and stellar orbits must be hierarchical, with the semimajor axis of a stable planetary orbit differing from the semimajor axis of the binary orbit by a factor of a few.

%%%%%%%%%% FIGURE 2 %%%%%%%%%%%%%%%%%%%%%%%%%%%%%%%%%%%%%%%%%%%%%%%%%%%%%%%%%%%
\begin{figure}
\epsscale{1.2}
\plotone{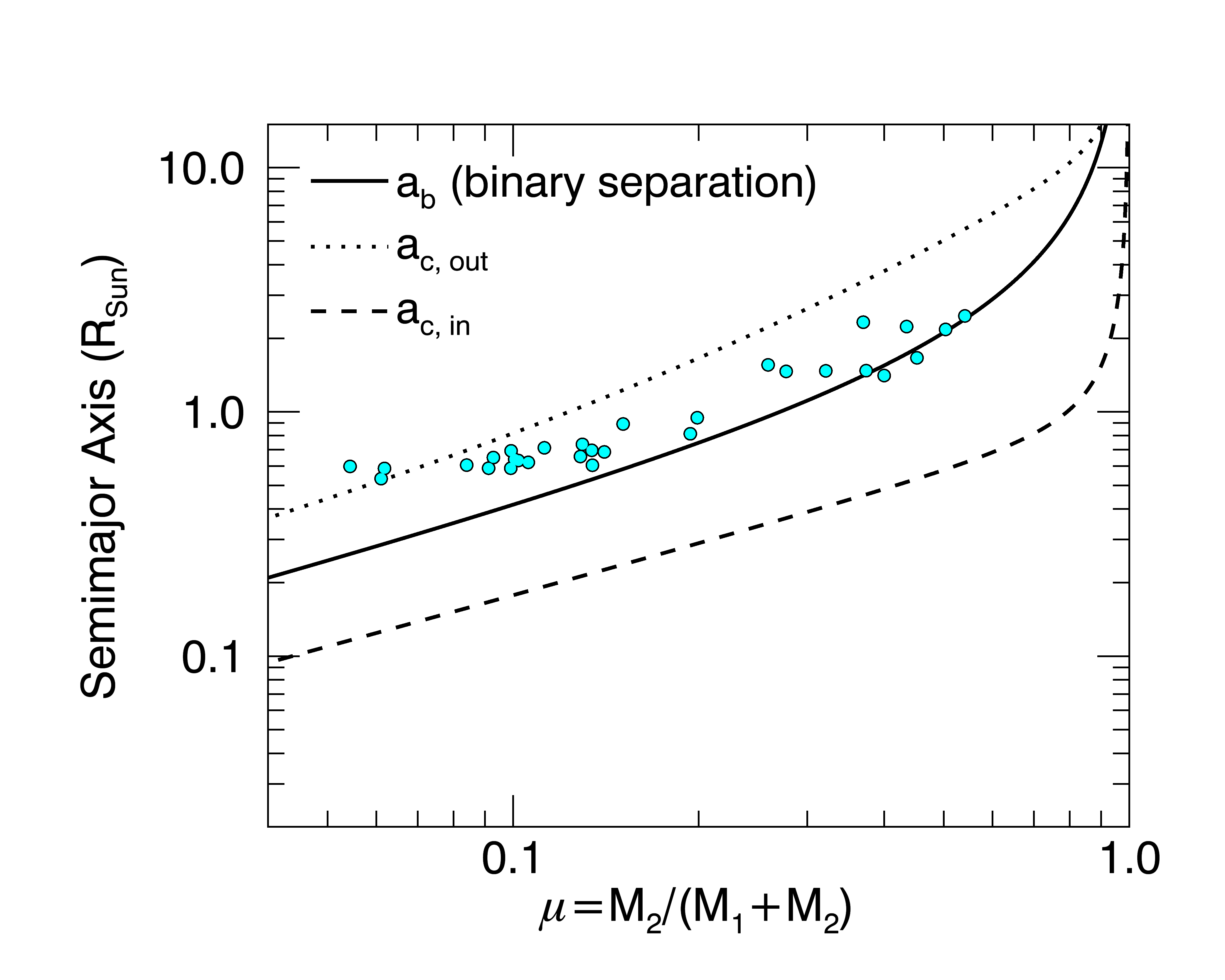}
\caption{Semimajor axes of the binary and of the planet as a function of $\mu$, for a fixed white dwarf mass of $M_1=0.8~\msun$.  The dotted curve is the critical semimajor axis of a planet in an outer orbit; the dashed line is the critical semimajor axis of a planet in an inner orbit around the white dwarf.  The cyan data points are the values of $a_b$ and $\mu$ for observed CVs from Zorotovic et al. (2011).  The mean white dwarf mass in the Zorotovic et al. sample is $0.8 \pm 0.2~\msun$. \label{fig:a_vs_mu}}
\end{figure}
%%%%%%%%%%%%%%%%%%%%%%%%%%%%%%%%%%%%%%%%%%%%%%%%%%%%%%%%%%%%%%%%%%%%%%%%%%%%%%%

When mass transfer occurs because the donor fills its Roche lobe, the binary's orbital separation is set by the condition that the radius of the donor, $R_2$, equals 
the size of its Roche lobe, $R_L$. Using
the function derived by  \citet{Eggleton_1983}, we have: 
\begin{equation}
\frac{a_b}{R_2} = \frac{0.6 q^{2/3} + \ln(1+q^{1/3})}{0.49 q^{2/3}},
\end{equation}
where $q = M_2/M_1$.  

In Figure \ref{fig:a_vs_mu}, we display the binary separation of XRBs 
as a function of $\mu$, for a fixed accretor mass of $M_1=0.8~\msun$,
corresponding to a white dwarf.   
We consider a case in which a main-sequence donor, with 
$R_2=M_2(R_\odot/M_\odot)$, fills its Roche lobe. 
The orbits are circular ($e=0$).   
The plot also shows the critical semimajor axes  
for  a single planet. We show both outer ($a_{c, \rm out}$; dotted line) 
and inner ($a_{c, \rm in}$; dashed line) critical orbits.  
Overplotted are data points from \citet{Zorotovic_2011} of binary separations for real CVs. There are no CVs on the upper right portion of the curve because mass transfer becomes unstable when the donor
is much more massive than the accretor.  The CVs depart from our curve on the lower left because the low-mass donors in these systems are degenerate and have larger radii than would be given by the extrapolation of the radius-mass relation for main-sequence stars. 

We note that the orbital separation can be significantly larger in mass
transfer binaries. The value of $a_b$ is generally $2$-$3$ times the radius 
of the donor, when the donor fills its Roche lobe. Thus, when the donor
is a subgiant or giant, $a_b$ is larger than in the main-sequence case
we have shown. In addition, mass transfer is often effected through winds.
The winds can be driven by radiation from the accretor or else may be 
generated by a highly evolved donor. In these cases, the separation is larger than it would be for Roche-lobe-filling systems.

Another factor that must be taken into account is the value of the mass ratio,
since large values are associated with mass transfer that is unstable on dynamical
time scales. For mass transfer mediated by winds, there is not a fixed analytic
 limit on the mass ratio, since larger mass ratios are allowed if much of the mass lost from the  donor exits the system carrying only a small amount of angular momentum per unit mass.   For mass transfer that occurs through the L1 Lagrange point, the limit depends on the evolutionary state of the donor, the fraction of the mass it loses that is ejected from the system, and the amount of
angular momentum carried by ejected mass. The critical mass ratio for stability may have values somewhat smaller than unity, although values  as large as $4$ can apply to main-sequence donors \citep[e.g.,][]{Ivanova_2015}. 
 
A planet that ventures too close to a white dwarf or other compact object would not be able to survive disruption by tidal forces.  The tidal disruption radius (the Roche limit) is  $R_T=1.16~R_\odot(\rho_p/\rho_\Earth)^{-1/3}(M_1/0.6~M_\odot)^{1/3}$, where $\rho_{p,\Earth}$ are the mass densities of the planet and of Earth,
respectively.  Tidal forces present a challenge for the survival of inner orbit (S-orbit) planets.

%%%%%%%%%% FIGURE 3 %%%%%%%%%%%%%%%%%%%%%%%%%%%%%%%%%%%%%%%%%%%%%%%%%%%%%%%%%%%
\begin{figure}
\epsscale{1.05}
\plotone{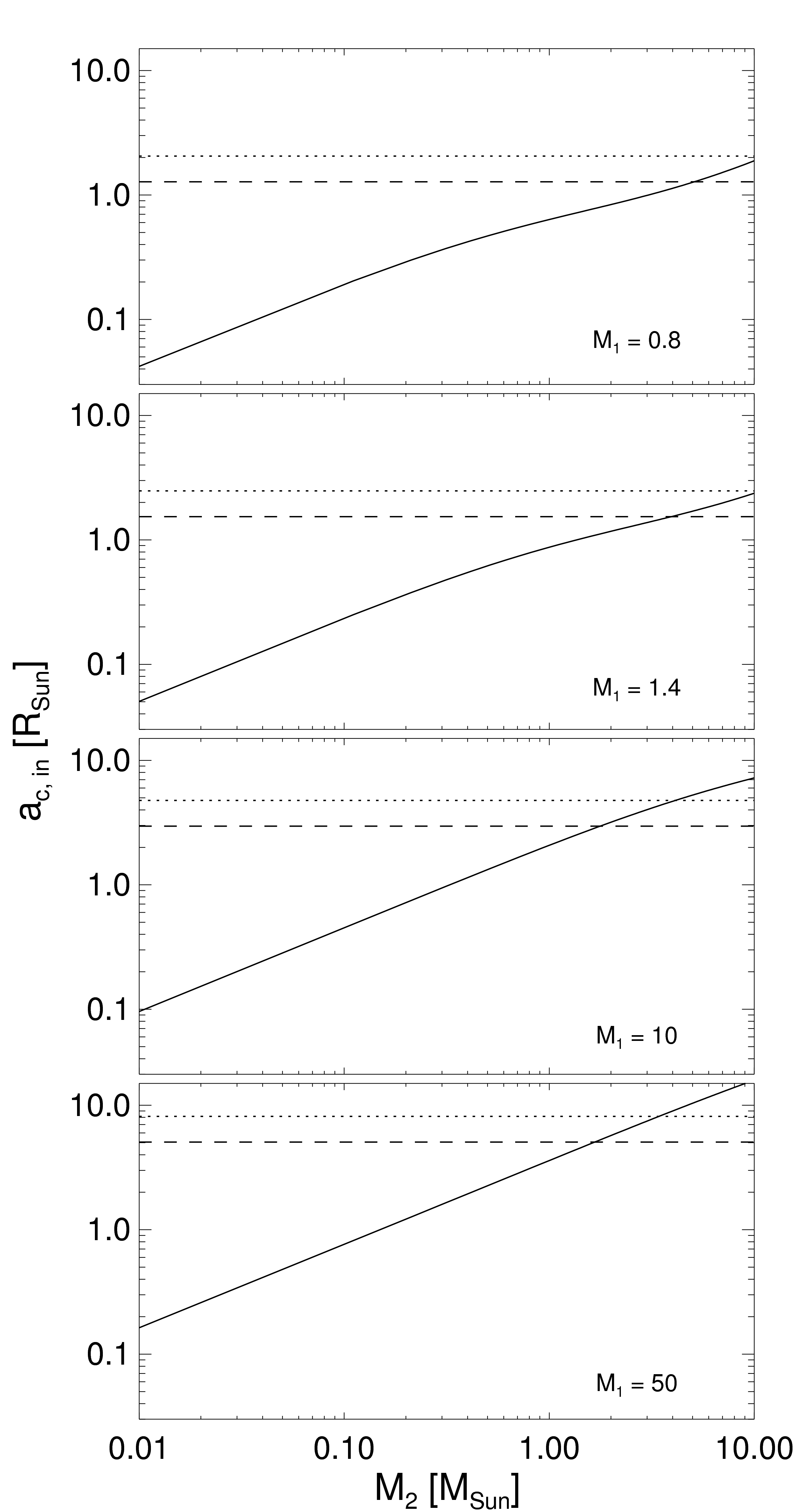}
\caption{Critical inner semimajor axis, $a_{c,\rm in}$, of a binary versus donor mass, $M_1$.  The results are plotted as solid curves for four different primary masses, $M_1 = 0.8$, 1.4, 10, and 50 \msun.  The dashed (dotted) line marks the location of the tidal radius, $R_T$, for a planet having the volume mass density of the Earth (Jupiter). \label{fig:tidal}}
\end{figure}

%%%%%%%%%%%%%%%%%%%%%%%%%%%%%%%%%%%%%%%%%%%%%%%%%%%%%%%%%%%%%%%%%%%%%%%%%%%%%%%

In Figure  \ref{fig:tidal}, we show the orbital parameters 
for several Roche-lobe filling binaries. 
Each panel shows the critical inner semimajor axis, $a_{c,\rm in}$, as a 
function of donor mass.  We show plots for four different values of the primary 
star's mass, $M_1=0.8$, 1.4, 10, and 50 \msun.
The horizontal dashed line marks the tidal radius for an 
Earth-density planet, ($\rho_\Earth\approx 5.5~\rm{g~cm}^{-3}; ~R_{T,\Earth}$, 
and a Jupiter-density planet,  ($\rho_{\rm J} \approx 0.24\rho_\Earth;~R_{T,\rm J}$).  Thus, for instance, a Jupiter-density planet in an XRB with a primary of mass $0.8~\msun$ (top panel) would be ripped apart for binary orbital separations less than $R_{T,\rm J}\approx 2.1R_\odot$.  An Earth-density planet in such a system could not survive orbital separations less than $R_{T,\Earth}\approx 1.3R_\odot$.

These plots illustrate several key points. First, as the formula for $R_T$
indicates, survival is more likely for planets of higher density. Second,
for accretors with masses in the range expected for white dwarfs and neutron stars, we do not expect to find stable S-type orbits if the donor star fills its Roche lobe. Thus, neither CVs nor low-mass X-ray binaries with neutron-star accretors are likely to have planets in stable S-type orbits. It is only for accretors of higher mass that we can find stable S-type orbits for binaries in which the donor is less massive than the accretor, ensuring the dynamical stability of mass transfer.  In fact, the discovery of a planet in an S-type orbit may place lower bounds on the accretor's mass, something very useful in the study of black holes.

The above results do not rule out S-type orbits in mass transfer binaries.
First, when the donor is larger, either a subgiant or giant, the orbital 
separation is larger even when the donor fills its Roche lobe. Second, when mass transfer occurs through winds, the  orbit can be even larger.  When the stellar orbits are wide, planets can orbit around either or both stars.

\section{Light Curves}\label{sec:light_curves}
To model the transit light curves, we need to assess the duration and the ingress/egress times of transits for a set of orbital parameters. A cartoon illustration of a transit in Figure \ref{fig:cartoon} shows that, for a non-grazing eclipse, the stellar and planetary disks touch at four contact times $t_{\rm I}$--$t_{\rm IV}$.  (For a grazing eclipse, the second and third contacts do not take place.) The transit duration $T$ and ingress/egress duration $\tau$ depend on the planet period, $P$, the radius of the primary star, $R_1$, the radius of the planet, $R_p$, the impact parameter, $b$, and the orbital inclination of the planet projected onto the plane of the sky, $i_p$.  For a circular orbit, the total duration, $T_{\rm tot} \equiv t_{\rm IV}-t_{\rm I}$, and full duration, $T_{\rm full}\equiv t_{\rm III}-t_{\rm II}$, are given by (e.g., Winn 2010),
\begin{equation}\label{eq:total}
T_{\rm tot} = \frac{P}{\pi}\sin^{-1}\left[\frac{R_1}{a_c} \frac{\sqrt{(1+k)^2 - b^2} }{\sin i_p}  \right],
\end{equation}
\begin{equation}\label{eq:full}
T_{\rm full} = \frac{P}{\pi}\sin^{-1}\left[\frac{R_1}{a_c} \frac{\sqrt{(1-k)^2 - b^2} }{\sin i_p}  \right],
\end{equation}
where $k=R_p/R_1$. The ingress and egress durations, $\tau_{\rm ing}=t_{\rm II}-t_{\rm I}$ and $\tau_{\rm egr}=t_{\rm IV}-t_{\rm III}$ are usually unequal for eccentric orbits, though the difference is minor in practice.  Here, since we consider only circular orbits, we take $\tau\equiv\tau_{\rm ing}=\tau_{\rm egr}$, and thus
\begin{equation}\label{eq:tau}
\tau=\frac{1}{2}(T_{\rm tot} - T_{\rm full}).
\end{equation}
In many XRBs, the source of  X-rays is a small region. It could for example be the inner accretion disk of a black hole or of a neutron star. In such cases the duration of the eclipse is determined by the amount of time it takes for the planet to cross this very small region.
In the case where the primary is a white dwarf of mass $M_1$, the size of the X-ray-emitting region can be close to the size of the white dwarf \citep[e.g.,][]{Nucita_2009}, $R_1$, which is determined by the white dwarf mass-radius relation: $R_1\approx 0.01(M_1/\msun)^{-1/3} R_\odot$.

%%%%%%%%%% FIGURE 4 %%%%%%%%%%%%%%%%%%%%%%%%%%%%%%%%%%%%%%%%%%%%%%%%%%%%%%%%%%%
\begin{figure}
\epsscale{1.2}
\plotone{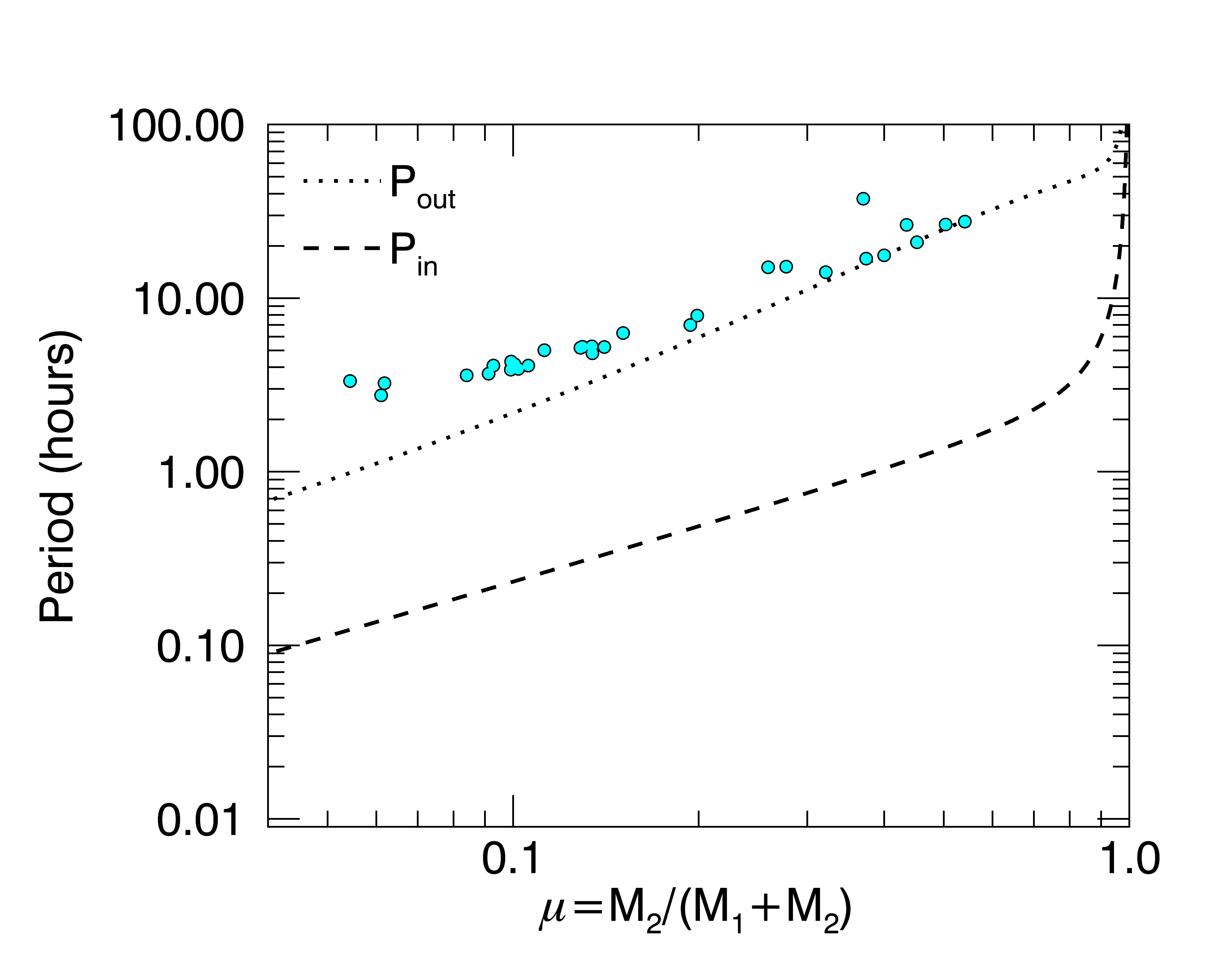}
\caption{Orbital period of a planet as a function of $\mu$, for a fixed white dwarf mass of $M_1=0.8~\msun$.  The dotted (dashed) curve represents the period for an outer (inner) orbit. The cyan points are the shortest possible periods of {\sl potential} planets in circumbinary orbits around the Zorotovic et al. (2011) sample of CVs, assuming outer critical semimajor axis orbits (see equation \ref{eq:outer}). \label{fig:period}}
\end{figure}
%%%%%%%%%%%%%%%%%%%%%%%%%%%%%%%%%%%%%%%%%%%%%%%%%%%%%%%%%%%%%%%%%%%%%%%%%%%%%%%

The orbital period of the planet is simply derived using Kepler's third law,
\begin{equation}
P^2 = \frac{4\pi^2 a_c^3}{G(M_\star + M_p)},
\end{equation}
where $M_p$ is the planet mass, and $a_c$ is determined from equations (\ref{eq:outer}) or (\ref{eq:inner}) for outer or inner orbits, respectively.  For inner orbits around the primary star,  $M_\star$ would refer to the mass of the primary star alone.  For outer orbits, however, the planet circles the combined mass of the primary and donor star, so $M_\star = M_1 + M_2$.  We consider Earth-mass and Jupiter-mass planets.  Figure \ref{fig:period} shows the planetary period for outer and inner orbits as a function of $\mu$ for a CV in which the primary is a 0.8 \msun~white dwarf.  Because $\mu$ varies and $M_1$ is fixed, this is equivalent to plotting the period for a range of donor masses, $M_2 = \mu M_1(1-\mu)$.  Since $M_\star\gg M_p$, the periods of Earth- and Jupiter-mass planets are essentially equal.  Recall that we have derived the period for values of the \emph{critical} semimajor axis, within which planetary orbits are unstable.  But orbits greater than $a_c$ are also permissible, and therefore the periods we calculate should be considered lower limits for a given  $M_\star$.  In Figure \ref{fig:period}, we overplot the periods of potential planets in circumbinary (CB) orbits around the \citet{Zorotovic_2011} sample of CVs.  Using the measured masses of the binary members to define $\mu$, we calculate the periods assuming the planets are in stable orbits at the critical semimajor axis, $a_{c,\rm out}$.  These calculations suggest that the periods of CB planets around XRBs may be as short as a few to tens of hours.

For completeness, we plot the period for inner orbits in Figure \ref{fig:period} as well, though based on the analysis of the tidal disruption radius in \S\ref{sec:stability}, such orbits may be physically implausible in CVs and other low-mass XRBs.

%%%%%%%%%%% FIGURE 5 %%%%%%%%%%%%%%%%%%%%%%%%%%%%%%%%%%%%%%%%%%%%%%%%%%%%%%%%%%%%%%
\begin{figure}
\epsscale{1.2}
\plotone{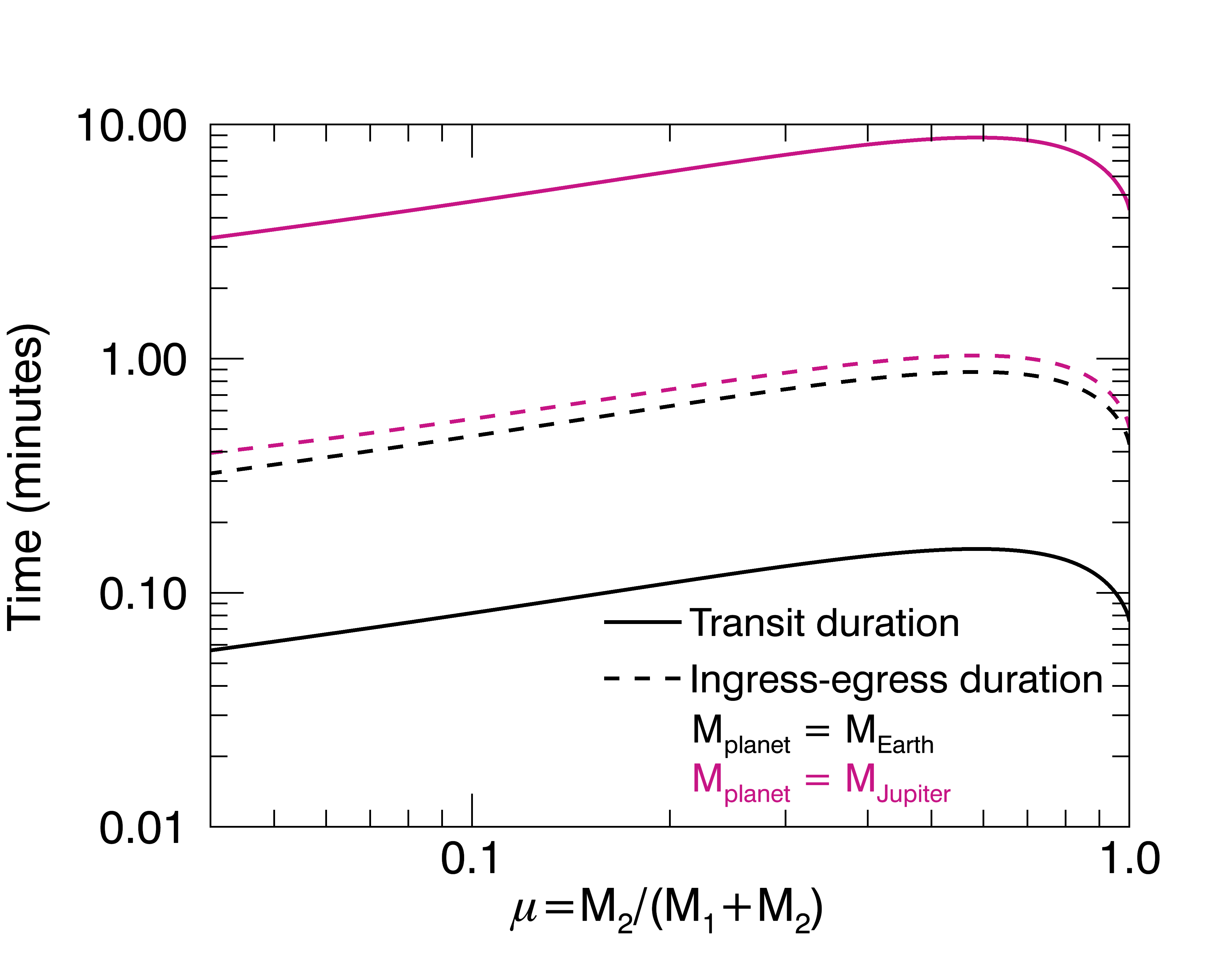}
\caption{Transit and ingress/egress duration of planets in the smallest possible circumbinary orbits (i.e., at the stability limit) around a binary consisting of a white dwarf with mass $0.8~\msun$.  The black and magenta lines represent the durations for an Earth-size and Jupiter-size planet, respectively. Note that ingress and egress dominate for the Earth-size planet. \label{fig:time_outer}}
\end{figure}
%%%%%%%%%%%%%%%%%%%%%%%%%%%%%%%%%%%%%%%%%%%%%%%%%%%%%%%%%%%%%%%%%%%%%%%%%%%%%%%%%%%

Figure \ref{fig:time_outer} plots the full duration and ingress/egress transit times versus $\mu$ for both Earth- and Jupiter-mass planets orbiting outside the binary. Again, we show the scenario where the primary mass is fixed to that of a 0.8 \msun~white dwarf.   And we focus our attention on simplified cases where the planet transits directly across the equator of the primary, i.e., $i_p=90\degr$ and $b=0$. 
The full duration and ingress/egress times displayed in Figure \ref{fig:time_outer} are lower limits for the scenarios considered; that is, these are the minimum transit times expected for CB systems in which the planets are near the stability limit.  If they exist, planets in XRB systems may certainly populate orbits beyond the stability limit.  Yet in the currently known CB systems discovered with \emph{Kepler}---if they can provide us with any direction---the planets are mostly close to the stability limit \citep{Welsh_2015, Li_2016}. 
The general trend is for both transit and ingress/egress durations to increase with binary mass ratio, $\mu$.  This is because as either the primary or donor mass increases, the orbit at which a planet is stable is pushed further outward, corresponding to an increase in the orbital period.

Small planets are expected to have short transit durations, on the order of a few seconds, and comparatively longer ingress/egress times, because their sizes are comparable to the sizes of white dwarfs ($\sim R_\Earth$). Due to their similar sizes, a small planet will spend relatively little time fully covering the stellar disk and a longer time crossing the edge of the stellar disk (ingress/egress).  The inverse is true for larger, Jupiter-like planets, which have long transit durations, on the order of several minutes, and comparatively shorter ingress/egress times.  Jupiter, for instance, is about 8.5 times larger than the 0.01 $R_\odot$ white dwarf in the HT Cas binary observed by \citet{Horne_1991} and \citet{Nucita_2009}.  Because white dwarfs (and other compact objects) are so small, the ingress/egress times of both small and large planets should be comparable.

Simulated transit light curves for Earth- and Jupiter-mass planets in CB (P-type) orbits around CVs with different mass ratios $\mu$ are shown in Figures \ref{fig:transit1} and \ref{fig:transit2}.  The orbits are circular and have no inclination.  In Figure \ref{fig:transit1}, the white dwarf mass varies ($M_1=0.6$, 0.8, 1, and 1.2 \msun) in each of the panels, while the donor mass remains fixed at $M_2 = 0.1~\msun$.  In Figure \ref{fig:transit2}, the white dwarf mass is fixed at $M_1 = 0.8$ \msun~in each of the panels, while the donor mass varies ($M_2 = 0.05$, 0.1, 0.5, and 1 \msun). In each plot in Figures \ref{fig:transit1} and \ref{fig:transit2}, the critical semimajor axis of the planet, (which depends on the mass ratio of the binary), sets the orbital period and therefore the transit duration.

\section{Discussion}\label{sec:discussion}
In this section we give examples of planetary transit light curves based on simple theoretical considerations.  
Our theoretical examples illustrate key features, but are not designed to capture the full complexity of real X-ray light curves, which may be affected by geometrical effects and time variability. Fortunately, there are many published examples of X-ray eclipses by stars, detected in spite of all of the complicating factors \citep[e.g.][]{VanTeeseling_1997, Ramsay_2001, Porquet_2005, Cherepashchuk_2009, Nucita_2009, Ponti_2017, Kennedy_2017}.  
In most cases, a dip in the X-ray light curve that is shorter in duration and/or more shallow than the stellar eclipse would also be detectable. These systems therefore provide examples in which planetary transits can either be discovered, or else the size and orbit of any transiting planet can be constrained if no signal is detected.

\subsection{Example light curves}
\begin{figure}
\epsscale{1.}
\plotone{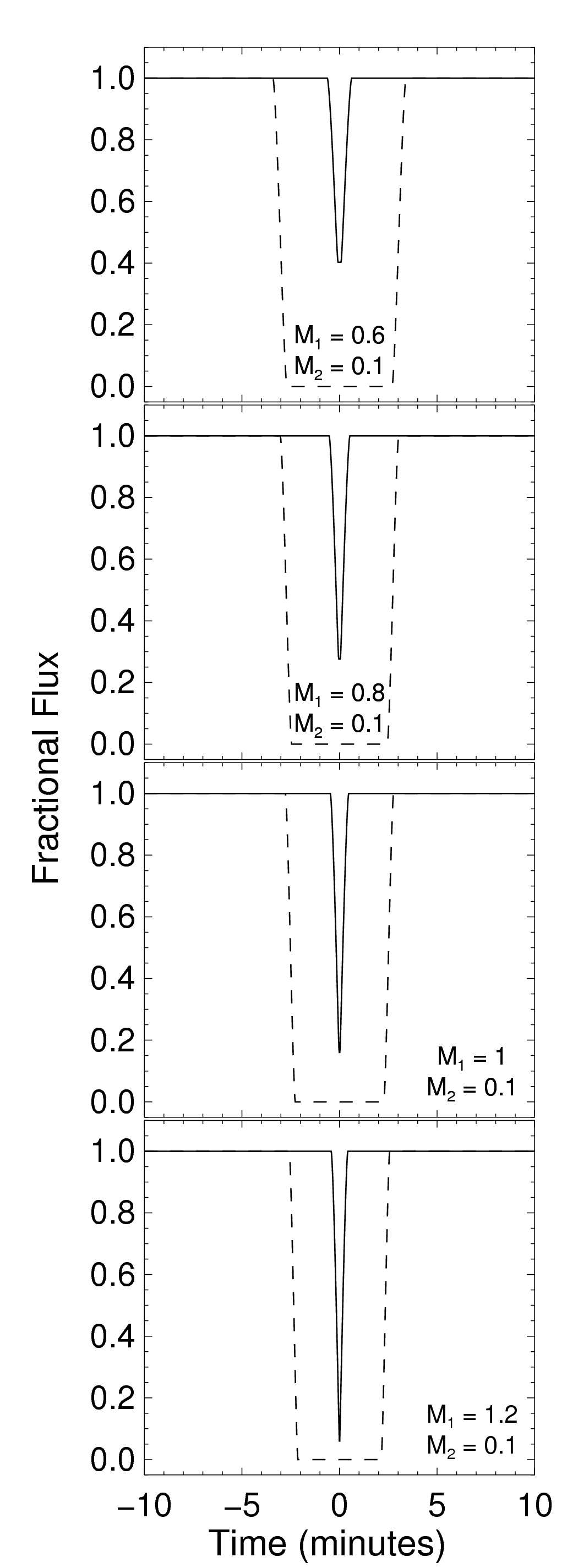}
\caption{Transit light curves for circumbinary planets, including an Earth-mass planet (solid lines) and a Jupiter-mass planet (dashed lines), for different values of $\mu$.  The white dwarf mass varies ($M_1 = 0.6$, 0.8, 1, and 1.2 \msun) in each of the panels, while the donor mass remains fixed at $M_2 = 0.1~\msun$.\label{fig:transit1}}
\end{figure}

In this section, we discuss the signatures we might expect in the light curves of CB planets in mass-transfer binary systems. 
In Figure \ref{fig:transit1} we plot light curves for a set of CVs in which the mass of the donor stays fixed at 0.1 \msun, and the white dwarf mass increases from 0.6 \msun~to near the Chandrasekhar limit at 1.2 \msun.  As the white dwarf mass increases, $\mu$ decreases, and therefore the transit duration becomes increasingly short.  The transit depth, $\delta=(R_p/R_1)^2$, for Earth-like planets gets deeper as $M_1$ increases, while Jupiter-like planets entirely eclipse the white dwarf for all mass ratios.  The former effect is due to the decrease in white dwarf size with increasing mass, according to the relationship $R_1\approx 0.01(M_1/\msun)^{-1/3} R_\odot$.  As the size of the white dwarf decreases---assuming the size of the X-ray emitting region in the CV is comparable in size to the white dwarf---an Earth-like (or larger) planet is capable of blocking out increasingly more of the flux, giving rise to deeper transit depths.  For $M_1=0.6$, an Earth-like planet would block $\delta \approx 0.4=40\%$ of the X-ray flux for a few seconds, while for $M_1=1.2$, it would block roughly 95\% of the flux for less time.

\begin{figure}
\epsscale{1.}
\plotone{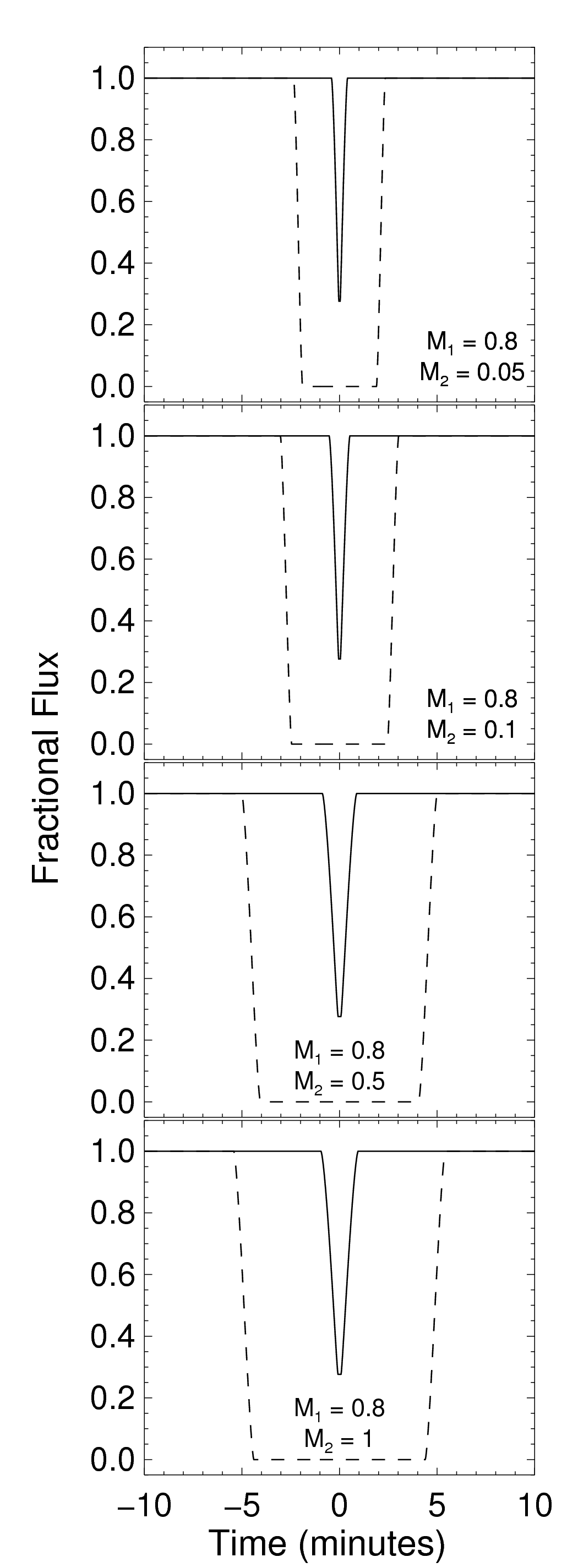}
\caption{Transit light curves for circumbinary planets, including an Earth-mass planet (solid lines) and a Jupiter-mass planet (dashed lines), for different values of $\mu$.  The white dwarf mass is fixed at $M_1 = 0.8$ \msun~in each of the panels, while the donor mass varies ($M_2 = 0.05$, 0.1, 0.5, and 1 \msun).\label{fig:transit2}}
\end{figure}

If we now consider CVs in which the white dwarf mass is fixed at a typical value of 0.8 \msun~and the donor mass increases, we find that the transit durations become increasingly longer, most noticeable for large, Jupiter-like planets (Figure \ref{fig:transit2}).  For a CV in which $M_1 = 0.8~\msun$ and $M_2=0.05~\msun$, a Jupiter-like planet would eclipse the white dwarf for roughly 4 minutes.  If instead, $M_2 = \msun$, a Jupiter would cover the white dwarf for roughly 9 minutes.  This trend is due to the increasing $a_c$, which increases with the mass ratio of the binary.  For smaller planets, the transit depth $\delta\approx 0.68$ remains fixed for all mass ratios; that is, roughly 68\% of the X-ray flux would be blocked out by an Earth-like planet orbiting a CV with a 0.8 \msun~white dwarf.

We note that for small, Earth-like planets, the duration over which the transit is deepest can be very short, lasting only a few seconds (see Figure \ref{fig:time_outer}).  This might raise doubts as to the possibility of detecting such small planets around XRBs.  However, it should be kept in mind that the ingress/egress durations of small planets are longer, around $\sim 30$--45 seconds.  Thus, the entire time scale of such a transit ($T+\tau_{\rm ing}+\tau_{\rm egr}$) will be $\gtrsim 1$ minute, an interval over which there could plausibly be detectable effects.

%%%%%%%%%%% FIGURE 6 %%%%%%%%%%%%%%%%%%%%%%%%%%%%%%%%%%%%%%%%%%%%%%%%%%%%%%%%%%%%%%
%%%%%%%%%%%%%%%%%%%%%%%%%%%%%%%%%%%%%%%%%%%%%%%%%%%%%%%%%%%%%%%%%%%%%%%%%%%%%%%%%%%

%%%%%%%%%%% FIGURE 7 %%%%%%%%%%%%%%%%%%%%%%%%%%%%%%%%%%%%%%%%%%%%%%%%%%%%%%%%%%%%%%
%%%%%%%%%%%%%%%%%%%%%%%%%%%%%%%%%%%%%%%%%%%%%%%%%%%%%%%%%%%%%%%%%%%%%%%%%%%%%%%%%%%

\subsection{Probability of Transits}\label{sec:trans_prob}
To estimate transit probabilities for circumbinary planets orbiting XRB systems, we follow \citet{Li_2016}, who provide analytical expressions for calculating transit probabilities involving eclipsing stellar binaries.  The transit criterion can be separated into two parts.  First, we calculate the geometric condition called transitability, in which the planet orbit intersects the stellar binary orbit.  Transitability does not guarantee actual transits on every crossing of the planet past the binary orbit, due to the relative motion of the three objects \citep{Martin_2015}.  Thus, our second step is to determine the planet transit probability, given that it intersects the binary orbit.  We  provide the following summary of the probability calculations and refer the reader to \citet{Li_2016} for the full details.

Both the mutual inclination between the planetary and binary orbits, $\delta i$, and orbital precession can affect the transitability.  Precession is described by time variation in the codependent quantities $\delta i$ and $\Omega$, the longitude of ascending node of the planet.  \citet{Li_2016} derive the probability for a planet to cross the binary orbit, $P_{\rm cr}$, as
\begin{equation}\label{eq:tran}
P_{\rm cr} = \frac{\rm{min}[\delta\Omega, 2\pi]}{2\pi},
\end{equation}
where $\delta\Omega = \Delta\Omega + \dot{\Omega}T_{\rm obs}$ is the total range of ascending node longitudes for the primary star, $\Omega$, during the total observation time, $T_{\rm obs}$. The total range of $\Omega$ over which the planet can cross the binary orbit is $\Delta\Omega$, which depends on the true anomaly of the planet, the planetary and binary semimajor axes, the size of the star being transited, the mutual inclination angle $\delta i$, and the inclination angle of the binary orbit, $i_b$ \citep[equations $10$-$12$ in][]{Li_2016}.  We slightly modify equations (11) and (12) in Li et al. to include the size of the transiting planet, $R_p$.  Li et al. do not include $R_p$ since, for the scenarios they consider, $R_p\ll R_1$.  However, for XRB systems, the transiting planet may have a size comparable to or larger than the size of the X-ray emitting region and cannot be ignored.  Thus, in Li et al.'s equations (11) and (12), we replace every occurrence of $R_1$ with $(R_1+R_p)$ (Gongjie Li, priv. communication).

%%%%%%%%%%% FIGURE 8 %%%%%%%%%%%%%%%%%%%%%%%%%%%%%%%%%%%%%%%%%%%%%%%%%%%%%%%%%%%%%%
\begin{figure}
\epsscale{1.2}
\plotone{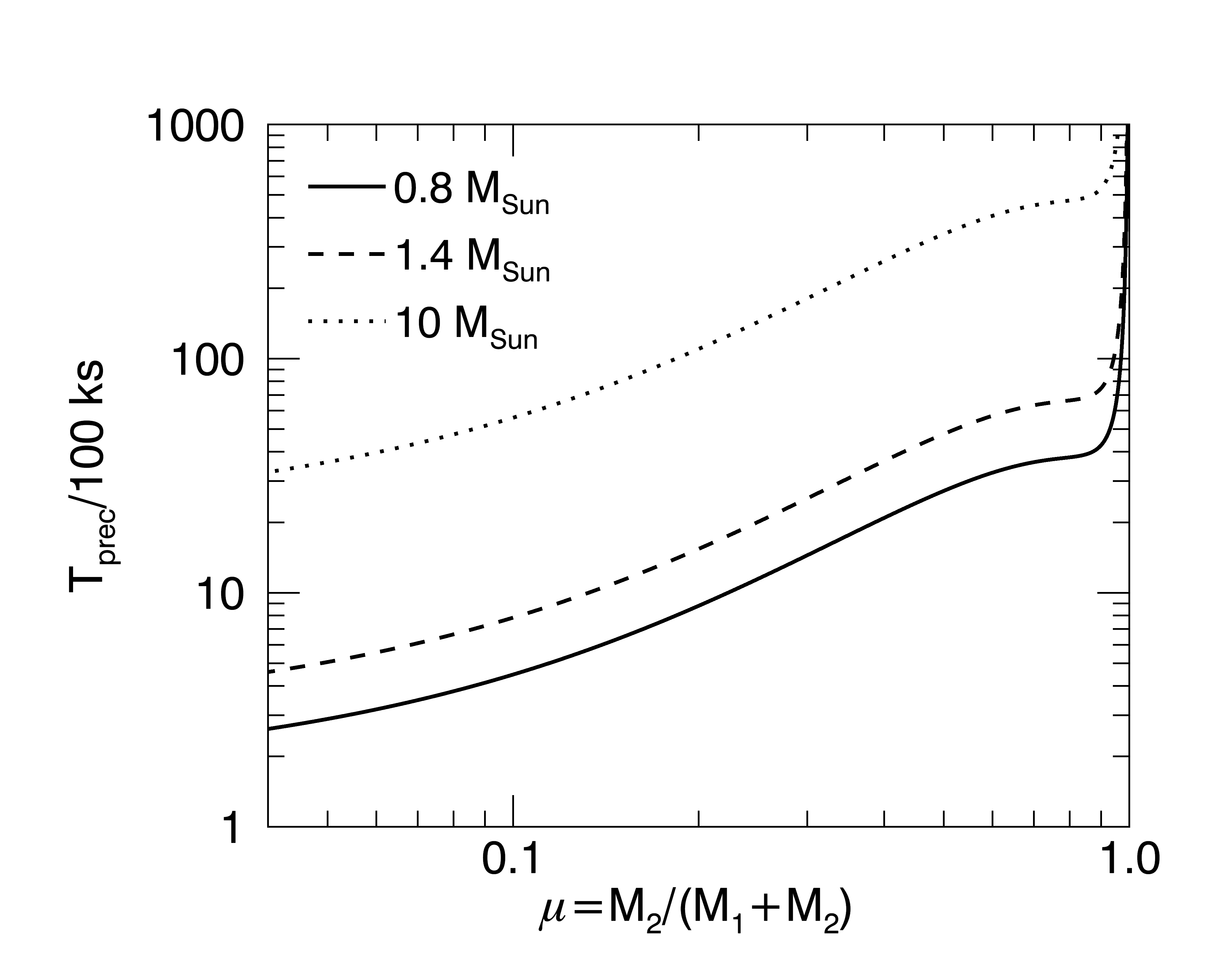}
\caption{Planetary precession period, $T_{\rm prec}$, as a function of binary mass ratio.  Examples are provided for three different primary masses: 0.8, 1.4, and 10 \msun.  $T_{\rm prec}$ is given in terms of a total observation time of $T_{\rm obs} =100$ ks.  \label{fig:prec}}
\end{figure}
%%%%%%%%%%%%%%%%%%%%%%%%%%%%%%%%%%%%%%%%%%%%%%%%%%%%%%%%%%%%%%%%%%%%%%%%%%%%%%%%%%%

The time rate of change of $\Omega$ due to precession is related to the orbital precession timescale of the planetary orbit by \citep{Schneider_1994}
\begin{equation}\label{eq:tprec}
\begin{split}
T_{\rm prec} &= \frac{2\pi}{|\dot\Omega|} \\
   &= P_p \frac{4}{3\cos \delta i} \left(\frac{a_p}{a_b\mu} \right)^2 \frac{m_2}{m_1}, 
\end{split}
\end{equation}
where $P_p$ is the planetary period, and $a_p$ is the semi-major axis of the planet.  In Figure \ref{fig:prec}, we plot $T_{\rm prec}$ as a function of $\mu$ for three scenarios, $m_1=0.8$, $1.4$, and $10$ \msun, for the coplanar case $\delta i = 0\degr$. Since the stellar masses used to calculate $P_p$ are much greater than the planet masses, the periods of Earth- and Jupiter-mass planets are indistinguishable in the plot.  
The precession time is given in terms of an observing time of $T_{\rm obs} = 100$ ks (see below), and the figure demonstrates that for a typical observing window, the precession periods of the considered circumbinary systems are comparatively long, $T_{\rm prec}>T_{\rm obs}$.  This confirms that precession should be taken into account for calculations of the XRB transit probability.

Next, we calculate the probability of transit, $P_{\rm tran}$, given that the planet and binary orbits intersect.  \citet{Li_2016} show that $P_{\rm tran}$ depends on the ratio of the relative displacement of the planet and star with respect to the observer, $dl$, and the projected width of the stellar orbit.  The relative displacements are 
\begin{equation}
\begin{split}
dl_1 &= t_{\rm tran} | v_p \cos\delta i - \frac{2v_1}{\pi}| + R_1 + R_p \\
dl_2 &= t_{\rm tran} | v_p \cos\delta i + \frac{2v_1}{\pi}| + 2(R_1+R_p), 
\end{split}
\end{equation}
where $v_{p,1}$ are the orbital velocities of the planet and primary star, and $t_{\rm tran}= \rm{min}[ (\pi (R_1+P_p))/(2v_p \sin\delta i), T_{\rm obs}]$ is the time it takes for the planet to cross the stellar orbit.  When the planet and primary star are both closer to the observer with respect to the center of mass, the relative displacement is $dl = dl_1$.  When the star is on the other side of the center of mass, the planet and star move in opposite directions, and $dl=dl_2$.  

%%%%%%%%%%% FIGURE 9 %%%%%%%%%%%%%%%%%%%%%%%%%%%%%%%%%%%%%%%%%%%%%%%%%%%%%%%%%%%%%%
\begin{figure*}[ht]
\centering
\includegraphics[width=5.7in]{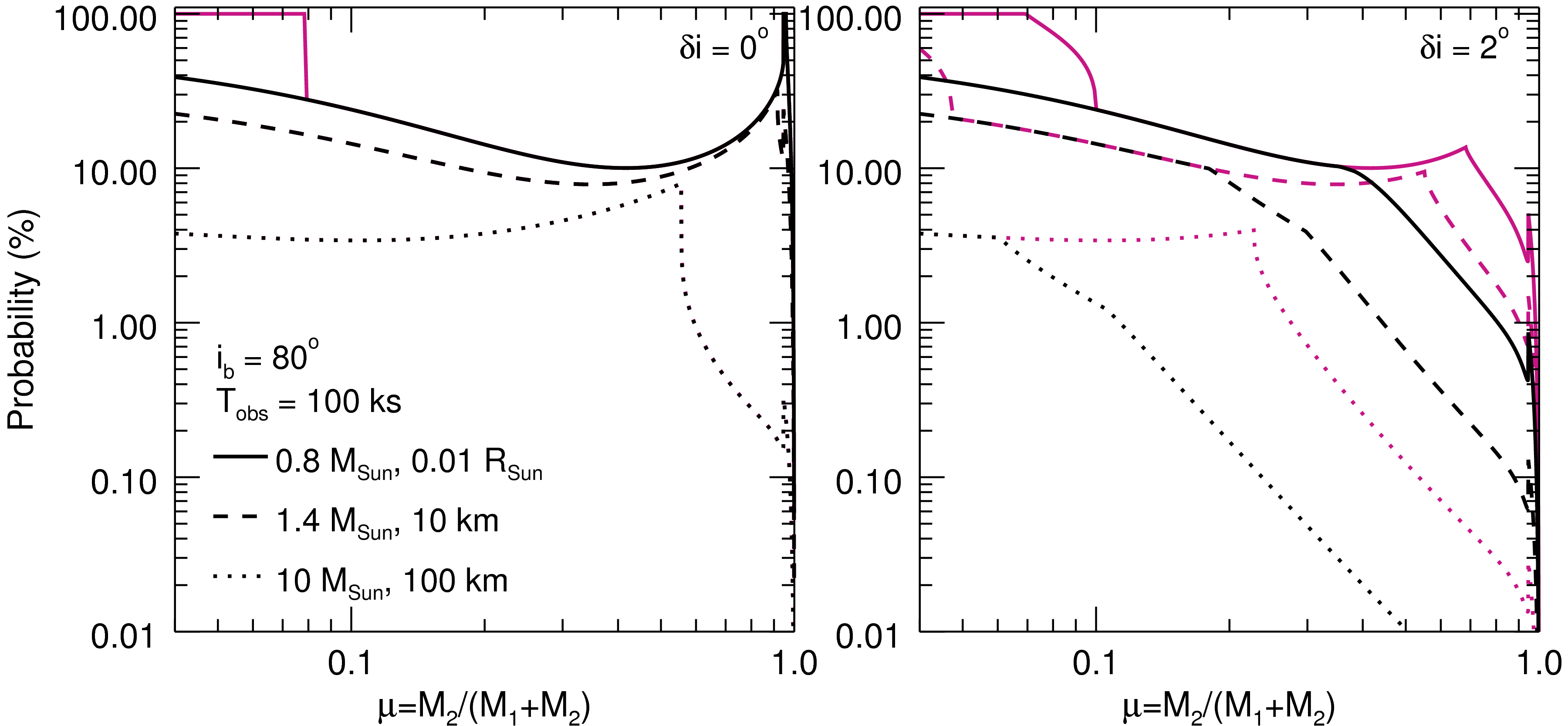}
\caption{Transit probability versus binary mass ratio, $\mu$, for planet circumbinary orbits around XRBs with primary masses of $0.8$ \msun~(solid lines), $1.4$ \msun~(dashed lines), or  $10$ \msun~(dotted lines).  The example systems, representing white dwarf, neutron star, and black hole binaries, also have different sizes for the X-ray emitting region: 0.01 \rsun, 10 km, and 100 km, respectively.  The black and magenta lines represent the probability calculations for an Earth-like and Jupiter-like planet, respectively.  A binary inclination angle of $i_b=80\degr$ and a total observation time of $T_{\rm obs} = 100$ ks are assumed.  The left panel assumes a mutual inclination angle between the planetary and binary orbits of $\delta i = 0\degr$.  The right panel assumes $\delta i = 2\degr$.    \label{fig:prob1}}
\end{figure*}
%%%%%%%%%%%%%%%%%%%%%%%%%%%%%%%%%%%%%%%%%%%%%%%%%%%%%%%%%%%%%%%%%%%%%%%%%%%%%%%%%%%

%%%%%%%%%%% FIGURE 10 %%%%%%%%%%%%%%%%%%%%%%%%%%%%%%%%%%%%%%%%%%%%%%%%%%%%%%%%%%%%%
\begin{figure*}
\centering
\includegraphics[width=5.7in]{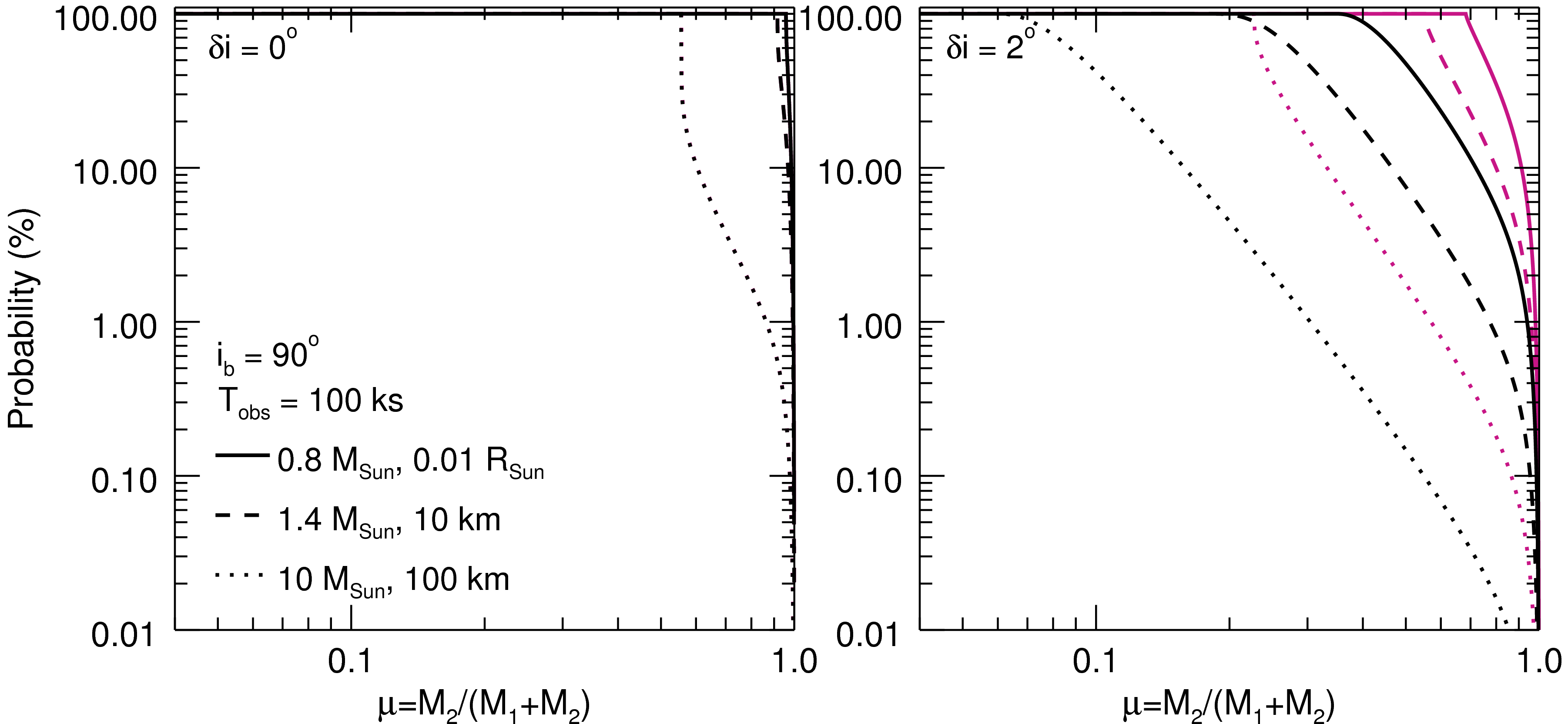}
\caption{Transit probability versus binary mass ratio, $\mu$, for a planet circumbinary orbits.  The curves represent the same types of systems shown in Figure \ref{fig:prob1}, accept we assume here that the binary orbits are edge-on with respect to the plane of the sky ($i_b=90\degr$).
   \label{fig:prob2}}
\end{figure*}
%%%%%%%%%%%%%%%%%%%%%%%%%%%%%%%%%%%%%%%%%%%%%%%%%%%%%%%%%%%%%%%%%%%%%%%%%%%%%%%%%%%

With these definitions, \citet{Li_2016} derive the probability for the planet to transit the primary star at least once in a finite observing window, of length $T_{\rm obs}$, as
\begin{equation}\label{eq:prob}
P_{\rm tran} = P_{\rm cr}[1-(1-P_1)^n],
\end{equation}
where
\begin{equation}\label{eq:pinf}
P_1 = 
\begin{cases}
1    &   \text{if}~\frac{dl_1 + dl_2}{2} > 2a_b\mu, \\
\frac{dl_1 + dl_2}{4a_b\mu}      &  \text{otherwise},
\end{cases}
\end{equation}
and where $2a_b\mu$ is the  projected width of the binary orbit. The exponent in equation (\ref{eq:prob}), $n$, is the number of times the planet crosses the binary orbit during a finite observing window.  The number of crossings is given by a conditional expression that depends on $T_{\rm obs}$, the planetary orbital period $P_p$, and precession (equation 18 in Li et al.).

Figure \ref{fig:prob1} displays the transit probabilities as a function of mass ratio, $\mu$, for a planet orbiting three different binary types: $m_1 = 0.8$ \msun~and $R_1=0.01~\rsun$ (solid lines); $m_1 = 1.4$ \msun~and $R_1=10$ km (dashed lines); and $m_1 = 10$ \msun~and $R_1=100$ km (dotted lines).  We assume a total observing time of $T_{\rm obs} = 100$ ks.  In the left panel, probabilities are calculated for coplanar systems, i.e., $\delta i =0\degr$.  In the right panel, we assume a mutual inclination of $\delta i = 2\degr$.  This value is motivated by observations of circumbinary planets detected by \emph{Kepler}, which exhibit a mean value of $1\fdg 7$ \citep{Martin_2015, Li_2016}.  For both panels, we assume an orbital inclination for the binary of $i_b = 80\degr$.  Modeling of  HT Cas suggests that the systems has an inclination of $81\degr$ \citep{Horne_1991}.

Figure \ref{fig:prob1} shows that for fixed values of $\mu$, $P_{\rm tran}$ increases as the mass of the primary star and size of the X-ray emitting region decrease.  Thus, systems representing CVs, with $m_1 = 0.8$ \msun, have the highest transit probabilities, while systems representing $10$ \msun~blacks holes and their companions have the lowest.  When mutual inclination effects are included, the relative behavior of $P_{\rm tran}$ is more complex.  In this example, increasing $\delta i$ by only $2\degr$ causes $P_{\rm tran}$ to decrease for most realistic values of the mass ratio, $0.05\lesssim\mu\lesssim 0.6$.  However, increasing  $\delta i$ even more would lead to an increase in the transit probability.  This is due to the competing effects between $T_{\rm obs}$, $dl_{1,2}$, and $\Delta\Omega$, all of which have different functional dependencies on $\delta i$.  As an example, if HT Cas ($\mu=0.19$, $i_B\approx 80\degr$) has a circumbinary Earth- or Jupiter-like planet at the stability limit, the transit probability would be $\lesssim 20\%$.

In Figure  \ref{fig:prob2}, we display plots of the transit probability, assuming that the binary orbital plane is completely edge-on with respect to the observer, i.e., $i_b = 90\degr$.  In these highly optimistic scenarios, values of $P_{\rm tran}$ are much higher than those for the examples depicted in  Figure \ref{fig:prob1}, where $i_b = 80\degr$.  Again, we see that a small increase in the mutual inclination between the binary and planetary orbital planes tends to cause a decrease in $P_{\rm tran}$.  In both Figures \ref{fig:prob1} and \ref{fig:prob2}, this decrease in  $P_{\rm tran}$ is especially pronounced for Earth-like planets (black curves).

In general, our calculations show that the transit probabilities for potential circumbinary planets in XRB systems are favorable for detection.  Taking the right panel in Figure \ref{fig:prob1} to represent the most conservative and plausible scenarios ($\delta i = 2\degr$ and $i_{ib} = 80\degr$), planets orbiting CVs may have transit probabilities in the range $2.5\%\lesssim P_{\rm tran}\lesssim 40\%$.  Planets in neutron star and black hole systems may have transit probabilities of $0.9\%\lesssim P_{\rm tran}\lesssim 20\%$ and $0.1\%\lesssim P_{\rm tran}\lesssim 4\%$, respectively, depending on the mass ratio of the system.

As indicated in Figures \ref{fig:prob1}-\ref{fig:prob2} and in equation (\ref{eq:tran}), the transit probability depends on the total observation time, $T_{\rm obs}$.  
We note that $T_{\rm obs}$ need not be continuous.   Rather, we are concerned with the total duration of all of the observations of a given X-ray source for which archived data is available. The archived resources are vast, including, from the present generation of X-ray missions alone, hundreds of mega-seconds of data.  The distribution of exposure times within a given year provides a general guideline.  
For example, observing programs approved by Chandra during Cycle 20\footnote{The Chandra web page with statistics for proposals: http://cxc.harvard.edu/target\_lists/cycle19/cycle19\_peer\_results\_ stats.html.} included 171 targets with exposures of 16 ks or less, with 120  of these having exposure times between 8 ks and 16 ks. In addition, there were 160 targets with exposures having durations ranging from 16 ks to roughly 1 Ms.  Many systems of special interest have been targeted for long observations, favorable for the discovery of transits.  These include some individual X-ray binaries, such as IC~10~X-1, an eclipsing accreting black hole in a Local Group dwarf galaxy. They also include globular cluster such as M15, and external galaxies located within 10 Mpc, such as M51, M82, and M101.

Finally, we comment on our choices for $R_1$, in equation (\ref{eq:tran}) and Figures \ref{fig:prob1}--\ref{fig:prob2}, necessary to calculate $P_{\rm tran}$.  The sizes of X-ray emitting regions depend on the nature of the compact accretor and also on the details of the accretion and emission processes. For instance, for a white dwarf that has undergone a nova explosion, soft X-ray emission emanates from the surface of the star, and the effective radius of emission is comparable to its radius, typically on the order of $0.01\,\rsun$.  Several luminous supersoft X-ray sources (SSSs) in the Magellanic Clouds and Milky Way are thought to be quasisteady-burning white dwarfs, and their X-rays are emitted from an effective area comparable to that of a white dwarf. The SSS class also includes white dwarfs post-nova \citep[][]{Henze_2016}.  Neutron stars can emit from their surfaces or, in some states, from the regions around the magnetic poles \citep{Wang_2017}. Emission can also come from a somewhat larger accretion disk, which is hottest near the inner radius.  \citet{Urquhart_2016} have used long exposures taken of M51 by both Chandra and XMM-Newton to discover X-ray transits of two separate X-ray sources, likely to contain black hole accretors.  These particular systems can exhibit emission from an inner disk having dimensions equal to that of the last stable orbit, 10s to 100s of kilometers for stellar mass black holes, and scaling up for black holes of larger mass \citep[][]{Soria_2017}.  
 
In sum, if Earth- and Jupiter-size planets exist in circumbinary XRB systems, and if their orbits are close to the minimum values set by stability requirements, then the transit probabilities show promise for detection.  A planet orbiting a single star has a transit probability of $(R_1 + R_p)/a_p$, and it is often the case that $R_1\ll a_p$, corresponding to a small value for the transit probability.  In an XRB, however, the semimajor axis of a planet may be so small that---despite the emitting region in the XRB also being potentially very small (see \S\ref{sec:possibility})---the ratio determining the \emph{transitability} probability (equation \ref{eq:tran}), and therefore $P_{\rm tran}$, is comparatively much larger.  Other factors, such as inclination and precession, may also favor high transit probabilities.  

We reiterate that our calculations assume that planets are near the stability limit.  If planets happen to be in wider orbits, this would decrease the transit probability.  Although there would be slight modifications in our calculations, the transit probability would essentially fall off as the inverse of the planetary separation, $a_p$ (Gongjie Li, priv. communication).  Thus, in our example above for HT Cas, if a planet was in an orbit at a radius $N$ times the critically stable orbit, the transit probability would be $\lesssim (1/N)\times20\%$.

In the next section, we discuss conditions that influence the detectability of transiting planets.

\subsection{Detectability and Identifying Candidate Systems}
For any given XRB with transiting planets, the identification of transits  
depends on (1)~the baseline, i.e., the out-of-eclipse background-subtracted 
count rate of the source, $dc/dt$; (2)~the duration, $\Delta\, t$, of the occultation; 
and (3)~the depth of the occultation. The depth of the occultation depends on the relative surface areas of the source and planet, as well as on the inclination of their orbit. 
Let ${\cal A}(t)$ be the area of the source which is covered by
the planet at some time $t$ after the first contact has been made 
and before the last contact is made. The average value of the area covered is 
${\cal A}_{\rm cov} = \langle {\cal A}(t) \rangle$.

The number of counts ``missing'' during the transit is then
\begin{equation}\label{eq:counts}
C = \frac{dc}{dt} \, 
\Bigg\{{\rm min}\Bigg[\Bigg(\frac{{\cal A}_{\rm cov}}{\pi R_{\rm em}^2} \Bigg)^2 ,1\Bigg]\Bigg\}\, 
 \times \Delta\, t,
\end{equation}
where $R_{\rm em}$ is the size of the X-ray-emitting region.

Exact expressions for $\Delta\, t$,  which include the geometrical 
effects,  are given in \S~2.
Here we express $\Delta\, t$ simply as $2\, R_{\rm eff}/v$, where $v$ is the
orbital speed during the eclipse, and $R_{\rm eff}$ is an effective radius. 
\begin{equation}\label{eq:delta_t}
\Delta \, t = 17~{\rm min}\, \Bigg(\frac{R_{\rm eff}}{R_J}\Bigg)\,
\Bigg(\frac{a_b}{10\, R_\odot}\, \frac{M_\odot}{M_\star} \Bigg)^{\frac{1}{2}}, 
\end{equation}
where $M_\star$ is the total mass enclosed by the orbit, and $R_J$ is the radius of Jupiter.  
In the limit of a small source and/or a large planet, $R_{\rm eff}$ may 
be taken to be the radius of the planet, $R_p$. 
More generally, $R_{\rm eff} = f \times (R_{p}+R_{\rm em})$, 
where $f$ is a number (generally $f \lesssim 1$) which takes into 
account orientation and ingress/egress effects.

Equation (\ref{eq:counts}) shows that, for a given count rate, more counts are missed 
during a transit for
larger orbital separation and smaller total mass. These same factors
yield longer orbital periods, however, producing fewer opportunities to 
detect additional transits that could validate the interpretation
that a dip in the count rate was associated with a transit.
Thus, for any given system, there is an optimal compromise 
between larger $a_b$ and smaller $M_\star$, which promotes the
detection of a single passage, and smaller $a_b$ and larger $M_\star$, which may make it
possible to detect multiple transits.  

To set scales relevant for the analysis of archived data, 
we consider the distribution of background-subtracted count rates obtained
during observations with the {\sl Chandra} X-ray 
observatory \citep{Wang_2016}. There were 772,000 X-ray states were found, and 
roughly 1000 (6000, 21,000) of them exhibited count rates  
larger than 1 s$^{-1}$ (0.1 s$^{-1}$, 0.01 s$^{-1}$).\footnote{
Wang et al. (2016) analyzed all of the {\sl Chandra} data collected
during the years $\sim$ 1999-2014. 
Many sources were detected multiple times, so that the
total number of distinct {\sl states} observed is about $3$ times larger
than the total number of distinct sources.} 
Many of the X-ray sources producing these states are XRBs. 
With the {\sl XMM-Newton} observatory, 
 count rates of the same sources 
would generally be several times larger, depending on the observational
set up \citep{Snowden_2002}. However, the number of background counts per unit time would also
be larger. 

For the highest count rates, a total eclipse lasting a
minute would be readily detected;  for
the lowest of these count rates, 
the passage might have to last for tens of minutes
to assure the reliable detection of a single planetary passage.
Repeated transits could be detected, allowing better analysis of
the dips, if the total exposure time, generally comprised of 
a sum of exposure times from separate  observations, encompasses
multiple orbits.  Some regions of the sky contain  
specific individual XRBs of high interest 
(e.g., black hole candidates),
and have been imaged many times. In addition,  
dozens of galaxies have had total exposure times in the range of 1-2 days,
with some sources (e.g., in M31) having had more coverage.
It therefore seems well within
the realm of possibility to detect planetary transits. 

\subsection{Physical Systems}

Already-discovered eclipsing XRBs can be used to guide searches through
archival data for planetary transits for a number of reasons. 
First, these systems provide specific examples
of a compact source being obscured, in that each exhibits a measured eclipse depth 
and duration.
By studying the signal-to-noise and other characteristics of the eclipse,
we can determine how small and how fast-moving a 
planet orbiting the X-ray source
 could be and still be detected. Such feasibility studies will also inform
analyses of other XRBs which do not exhibit stellar eclipses. 

Second, because stellar and planetary orbits may be coplanar, 
an XRB in which there is a stellar eclipse may be more likely to
also exhibit  planetary eclipses, should that system host planets.
Thus, XRBs with stellar eclipses may be the most promising 
systems in which to search for planetary transits.
The period of planetary transits would be different from the 
binary period; the stability issues we discuss here would provide
useful guides. 

Finally, a stellar eclipse establishes that the X-ray source is compact.
For example, the long-lasting X-ray eclipse in the black-hole/Wolf-Rayet binary
X1 in the dwarf galaxy IC~10, illustrates that some black hole binaries do
have compact regions emitting distinctive X-rays. Similarly, several
eclipsing white dwarfs are known, where X-rays emanate from near the
surface of the white dwarf \citep[e.g.,][]{Nucita_2009}. 
These indicate that it could be productive to search 
for planets orbiting white dwarfs that have recently 
experienced novae. Post-nova,
many systems pass through a {\sl supersoft} phase in which X-rays 
with energies $< 1$~keV are emitted from near the white dwarf's surface.
Such searches may find planets in P-type orbits around close-binary novae
and in S-type orbits in symbiotic novae, which generally have 
wide-orbit giant donors. 

Equation (\ref{eq:counts}) allows an estimate of the count rate necessary to allow transits to be identified. Count rates that are large enough for this purpose can be provided by nearby XRBs or by distant yet bright XRBs. Consider a source with X-ray luminosity equal to $10^{39}$~erg~s$^{-1}$ at $5$~Mpc; this corresponds to  an ultraluminous XRB in an external galaxy. If the X-ray spectrum is a power law of index $1.7$, {\sl XMM-Newton} would collect $0.1$ counts per second (cps).\footnote{HEASARC PIMMS predicts  $0.1$~cps with {\sl XMM} PN THIN (0.4--10 keV) for an on-axis observation and $N_H = 10^{21}$~cm$^{-2}$. (https://heasarc.gsfc.nasa.gov/cgi-bin/Tools/w3pimms/w3pimms.pl.) This can be sufficient for the discovery of transiting planets.}  The same flux would be obtained from a source  of $10^{33}$~erg~s$^{-1}$ at $5$~kpc; this could correspond to  a quiescent low-mass X-ray binary (LMXB) in a Galactic  globular cluster, while the flux from a bright globular cluster LMXB would be $1000$ times larger. Alternatively, a very dim source with X-ray luminosity of  $10^{29}$~erg~s$^{-1}$ at $50$~pc, would provide the same flux.      

As discussed in \S\ref{sec:trans_prob}, the probability \ptran~that the orientation is favorable for transit detection declines with increasing orbital separation. If an observation
capable of detecting transits lasts over a time interval of one or more orbital periods, the probability remains equal to \ptran. Even if a sequence of shorter observations are spread out over a long interval, the probability is roughly equal to \ptran~as long as the total observation time is longer than an orbital period. On the other hand, when the orbital period is longer than the total observing time, then the likelihood that the time interval of an observation will include a transit depends on the fraction of the time the system is in transit, the transit duty cycle: $\sim R/(\pi\, a)$.

{The question of how the duty cycle affects detectability primarily hinges on the typical time duration of the X-ray observations relative to the orbital period. Many X-ray sources that are promising targets for transit searches have been observed for hundreds of kiloseconds, including various regions in M31 \citep[e.g.,][]{Vulic_2016}. The same is true of roughly 10 nearby galaxies, in which the observing footprint covers most of the galaxy, and with each galaxy 
typically containing dozens of bright sources \citep[e.g.,][]{Jordan_2004, Sell_2011}. Finally, Galactic globular clusters and many individual sources in our galaxy and the Magellanic Clouds have been targets of long-term monitoring by several different X-ray observatories.}

These exposures are long enough to ensure that transits occurring in systems with  orbital periods up to tens of hours have a probability \ptran~of being observed.  Planets in wider orbits could produce detectable transits---and, in fact, the transit duration would be longer, enhancing the signal. But the probability would be reduced as described above.  
   
We don't know the orbital period distribution of planets in X-ray binaries. 
However, among known exoplanets discovered via the transit method, roughly one in eight have orbital periods shorter than two days.\footnote{See exoplanets.eu} We can gain insight about possible X-ray binary planetary systems from observations and calculations of the evolution of planetary orbits during stellar evolution. Planets in XRB systems would likely have been part of the planetary system of one of the binary's original  component stars. Dynamical interactions during the binary's evolution would have ejected some planets and placed others into closer orbits.   Recently, low-mass objects have been discovered orbiting WDs.
For example, eclipses of a WD (SDSS J141126+200911)  at optical wavelengths identify an object interpreted as a brown dwarf in a 121.73 minute orbit, producing eclipses 125 s in duration \citep{Beuermann_2013}.  Such discoveries focus attention on the question of how such systems evolve.  Calculations of the tidal interactions between evolving stars and their low-mass companions, and calculations of the possible engulfment of the low-mass companions find that there should be a period gap in the post-evolution orbits, with brown dwarfs able to occupy orbits with periods smaller than $0.1$~day \citep{Nordhaus_2010}.  Planets may come to be in close orbits around WDs if they survive engulfment, or if their initial orbits were wide enough to avoid engulfment, but they were subsequently scattered into close orbits \citep{Nordhaus_2013}.

\subsection{Search for extraterrestrial intelligence}
An interesting application of our proposal to search for planets around XRBs is that the same technique could be used in the search for extraterrestrial intelligence (SETI), since if advanced civilizations exist, they may produce
 anomalies in astronomical systems.  One possible  anomaly could be caused by placing around a star a Dyson sphere, a surface designed to capture a star's energy and convert the starlight into far-infrared emission \citep{Dyson_1960}.  
It has also been suggested that smaller, multiple structures could be placed in orbit around a star \citep[e.g.,][]{Harrop_2010}.  Should they exist, some such 
artificial structures might caused detectable transits.  

Such artificial structures are  often discussed in conjunction with the possible energy needs of an advanced civilization, and because any convincing discovery of
them would provide evidence of exocivilizations. Here we make a possible connection to XRBs, which are (1) highly luminous and therefore potentially desirable as sources of energy, and also (2) distinctive and rare enough to be targeted for X-ray  observations by emerging technological civilizations.   

We therefore suggest that searching for transits around XRBs could---in addition to or instead of leading to evidence of natural planets---lead to evidence of artificial structures. While it is not clear that we could distinguish between natural and artificial structures, one could attempt to answer such questions as: Are the planetary orbits stable? Does the light curve, especially during ingress and egress, suggest an unusual shape?

We also make a particular connection to globular clusters, which have been suggested as environments within which advanced civilizations may be able to thrive because of (1) the absence of young stars and the attendant risks, such as supernovae; (2) the stability of orbits in the habitable zones of the low-mass stars comprising the primary cluster populations; and (3) the relatively high stellar density, which could facilitate interstellar travel, should technologically advanced civilizations inhabit globular clusters \citep{Di_Stefano_2016}.  XRBs are rare, but are more common in globular clusters than in the rest of the Galaxy \citep{Clark_1975, Di_Stefano_1994}. Furthermore, they could be significant sources of energy in an environment dominated by M dwarfs, while at the same time lying within a relatively small distance of any point in the cluster, due to the limited (generally $< 10$ pc) size of globular clusters.

\section{Summary}\label{sec:conclusion}

We have presented a novel idea for searching for exoplanets around mass 
transfer X-ray binaries.  For demonstration of concept, we have 
focused on CVs, in which the typical mass of the 
white dwarf is $\sim 0.8$ \msun~and the donor is a main-sequence star
or a brown dwarf. If a planet is associated with such a binary,
it is most likely to have a circumbinary (P-type) orbit. 
While, for a mass ratio of $0.5$, the closest possible orbit would exhibit 
transits of just under a minute for an Earth-size planet and about $10$ times
longer for a Jupiter-size planet, the transit durations would be 
larger in wider orbits. We have shown how these quantities scale for
accretors of higher mass (neutron stars and black holes) 
and for a range of mass ratios.   
If the binary has a wider orbit, then either the compact  
object or the donor star, may support a planet in an S-type orbit.  

We have demonstrated that a number of factors may favor high transit probabilities form potential circumbinary XRB planets.  If planets are in orbits near their critical semi-major axes, and if inclination and precession are important, then transit probabilities may range from roughly 0.1\% to 40\%, depending on the mass ratio of the binary and size of the X-ray emitting region.  
Given our probability calculations, we have also shown that transits of planets orbiting XRBs, including
symbiotic binaries, are detectable with present-day technology. 
Future observations with missions such as the proposed
{\sl Lynx} X-ray mission,\footnote{https://wwwastro.msfc.nasa.gov/lynx/} 
which may have $50$ times the sensitivity of {\sl Chandra}, will dramatically extend the opportunities for transit detection.   The discovery of a planetary transit in X-ray binaries, while not assured, would be of major scientific importance.  Any well-designed archival study will either discover planets or else will place limits on planetary orbits.  If and when the first planet is detected around a mass transfer binary, follow-up observations could be conducted with other telescopes  to constrain its atmospheric properties and to search for biosignatures.

\acknowledgements
From Nia Imara: I dedicate this paper to my grandmother, Carrie Antonia Lee, and to Lewis Bundy.  I thank Matthew Holman, Gongjie Li, and Jason Wright for discussions and comments that helped improve this paper.  This work is supported by the John Harvard Distinguished Scientist Fellowship.

\clearpage
%--- BIBLIOGRAPHY ---------------------------------------------------
\bibliography{paper}
%\bibliography{bibliography}

\end{document}